\documentclass{iopart}
\usepackage{epsfig}
\begin{document}

\title[Optimum measurement for unambiguously discriminating   
two mixed states]{Optimum measurement for unambiguously discriminating   
two mixed states: General considerations and special cases}

\author{Ulrike Herzog\dag\ and J\'anos A. Bergou\ddag}  

\address{\dag\ Institut f\"ur Physik,  Humboldt-Universit\"at zu
    Berlin, \\
    Newtonstrasse 15, D-12489 Berlin, Germany}

\address{\ddag\ Department of Physics, Hunter College, City   
University of New York, \\ 
695 Park Avenue, New York, NY 10021,   
USA}   
\begin{abstract}   
Based on our previous publication [U. Herzog and J. A. Bergou, Phys. Rev. A {\bf 71}, 050301(R)(2005)] 
we investigate the optimum measurement  
for the unambiguous discrimination of two mixed quantum states that occur
with given prior probabilities. 
Unambiguous discrimination of nonorthogonal states is possible in a probabilistic way, at the expense 
of a nonzero probability of inconclusive results, where the measurement fails. 
Along with a discussion of the general problem, we give an example illustrating our method of solution. 
We also provide general inequalities for the minimum achievable failure probability
and discuss in more detail the necessary conditions that must be fulfilled when its absolute lower bound, 
proportional to the fidelity of the states, can be reached. 
\end{abstract} 

\section{Introduction}

Quantum state discrimination \cite{springer} is a basic tool for quantum information and quantum communication tasks. 
In the standard problem, 
it is assumed that a quantum system is prepared, with certain prior probability, 
in a certain state chosen from a finite set of given states. A state discriminating measurement 
determines what the actual state of the system is from the set of possible states. According to 
the laws of quantum mechanics, perfect state discrimination yielding a correct
result in each single measurement is impossible when 
the given states are not mutually orthogonal. For this general case 
various measurement strategies have been developed that are optimized with respect to 
different critera. For example, the measurement for minimum-error 
discrimination \cite{helstrom} yields a definite result for the state of the system each time it is performed, 
but this result may be wrong and the probability of errors is as small as possible.
In unambiguous discrimination, on the other hand, errors are not allowed to occur, 
which is possible at the expense of allowing measurement outcomes, with a certain probability of occurence, 
that are inconclusive, i. e. that fail to give a definite answer. 

In a measurement for optimum unambiguous discrimination the failure 
probability is minimized. Such a scheme was first introduced for distinguishing 
between two pure states \cite{ivan}-\cite{jaeger}, 
and only in the past few years the interest focused on investigating 
unambiguous state discrimination 
involving also mixed states \cite{SBH}-\cite{BFH}.  
While for optimum unambiguous discrimination between 
a pure state and an arbitrary mixed state an explicit general result has been 
derived for the minimum failure probability \cite{BHH,BHH1},
the problem of optimally discriminating between two arbitrary mixed states is much 
more complicated and there does not exist an explicit compact solution comprising 
the most general case. However, a number of important results have been obtained. 
First, an overall lower bound has been found for the failure probability \cite{rudolph}, 
being later on generalized for distinguishing between more than two mixed states \cite{feng}. 
Moreover, theorems have been established  
that reduce the problem of unambiguous discrimination to a standardized form \cite{raynal}. 
Recently the necessary conditions have been derived that must be fulfilled when 
the failure probability saturates the overall lower bound \cite{HB3},
and explicit expressions have been provided for the optimum measurement operators 
in the case that the two mixed states belong to a 
special class \cite{raynal1}.
In addition, the optimum measurement has been determined 
for a number of special cases having various degrees of 
generality \cite{rudolph}, \cite{HB3}-\cite{BFH}.

In this contribution we give a brief summary of the main results presented in 
our previous publication \cite{HB3}, along  with a more detailed discussion of  
several issues. We also show how our method can 
be applied to actually solve an illustrative special example.

\section{Description of the optimization problem and remarks on the solution}

We start by recalling the underlying theoretical concepts for investigating
a measurement capable of unambiguous discrimination of arbitrary quantum states.
Any measurement for distinguishing between 
two mixed states, characterized by the density operators 
$\rho_1$ and $\rho_2$ 
and occurring with the prior probabilities $\eta_1$ and $\eta_2= 1-\eta_1$, respectively,     
can be formally described with the help of three positive detection operators   
$\Pi_0$, $\Pi_1$ and $\Pi_2$, where
\begin{equation}
\Pi_0 + \Pi_1 + \Pi_2 =I
\label{Pi}
\end{equation} 
with $I$ being the identity. These operators are defined in such 
a way that  
$\rm Tr(\rho\Pi_k)$ 
 with $k=1,2$ is the probability that a system   
prepared in a state $\rho$ is inferred to be in the state $\rho_k$,  
while $\rm Tr(\rho\Pi_0)$ is the probability that the     
measurement fails to give a definite answer. 
The measurement is a von Neumann measurement when all detection operators
are projectors, otherwise it is  
a generalized measurement based on a positive operator-valued measure (POVM).  
When the detection operators are known,  
schemes for realizing the measurement can be devised  
\cite{neumark}.        

In unambiguous discrimination errors do not to occur  
which is equivalent to \cite{springer}   
\begin{equation} 
\label{Pi1}
\rho_1 \Pi_2 = \rho_2 \Pi_1= 0. 
\end{equation}
The total probability that the measurement fails can then be written as 
\begin{eqnarray} 
Q & = & \eta_1 \rm Tr (\rho_1\Pi_0) + \eta_2 \rm Tr (\rho_2\Pi_0)\nonumber\\  
& = & 1- \eta_1{\rm Tr}({\rho}_1{\Pi}_1) - \eta_2{\rm Tr}({\rho}_2{\Pi}_2),
\label{Qfail}
\end{eqnarray} 
where  in the second line Eqs. (\ref{Pi}) and (\ref{Pi1}) have been used. 
In order to determine the measurement, we have to find explicit expressions for the detection operators.
For this purpose we introduce the spectral representations of the given density
operators,  
\begin{equation} 
\label{rho} 
\rho_1 = \sum_{l=1}^{d_1} r_l |r_l\rangle\langle r_l|, 
\qquad    
\rho_2 = \sum_{m=1}^{d_2} s_m |s_m\rangle\langle s_m|,                  
\end{equation}   
where $r_l, s_m\neq 0$, and $\langle r_l|r_{m}\rangle = \delta_{l,m}   
=\langle s_l|s_m\rangle$. 
Moreover,  we   
introduce the projection operators    
\begin{equation}   
\label{10}   
P_1 = \sum_{l=1}^{d_1} |r_l\rangle\langle r_l|,    
\qquad   
P_2 = \sum_{m=1}^{d_2} |s_m\rangle\langle s_m|.                
\end{equation}   
Now we decompose each eigenstate of $\rho_1$ into a component 
lying within the support of $\rho_2$, i. e. within the Hilbert space spanned by the 
eigenstates of $\rho_2$ belonging to nonzero 
eigenvalues, and a second component being 
perpendicular to it, 
$|r_l\rangle= P_2 |r_l\rangle + |r_l^{\perp}\rangle$.
By applying the Gram-Schmidt orthogonalisation procedure \cite{nielsen} we can 
construct a complete orthonormal basis $\{|v_i\rangle \}$    
in the subspace spanned by the non-normalized 
state vectors $|r_l^{\perp}\rangle$,
($l=1\ldots,d_1$), and we denote the projector onto this 
subspace by $P_{1\perp}$.
In an analogous way, after decomposing the eigenstates of $\rho_2$,
we determine
the projector $P_{2\perp}$. 
When $d_{1\perp}$ and $d_{2\perp}$ are the dimensionalities of the 
corresponding bases,  
these projectors take the form   
\begin{equation}   
\label{12}   
P_{1\perp} =  I - P_2 = \sum_{i=1}^{d_{1\perp}}|v_i\rangle\langle v_i|,\qquad
P_{2\perp} =  I-P_1 =\sum_{j=1}^{d_{2\perp}} |w_j\rangle\langle w_j|, 
\end{equation}   
where  $\langle v_i|v_j\rangle=\langle w_i|w_j\rangle = \delta_{i,j}$.  
Since by construction $\rho_2 |v_i\rangle = 0$ for $0\leq i \leq d_{1\perp}$ 
and  $\rho_1 |w_j\rangle = 0$ for $0\leq j \leq d_{2\perp}$,
the most general Ansatz for the detection operators $\Pi_1$ and  $\Pi_2$ 
can be written as 
\begin{equation}   
\label{detect1}   
\Pi_1    
=\sum_{i,j=1}^{d_{1\perp}} \alpha_{ij}|v_i\rangle\langle v_j|, \qquad
\Pi_2    
=\sum_{i,j=1}^{d_{2\perp}} \beta_{ij}|w_i\rangle\langle w_j|.  
\end{equation}   
Clearly, unambiguous discrimination is possible with a non-zero probability of success  
when the supports of the two density operators are not 
identical, since in this case at least one of the operators $\Pi_1$ and $\Pi_2$
does not vanish. On the other hand, the measurement result is always inconclusive when the 
supports cooincide, $P_1=P_2=I$, because in this case $\Pi_1=\Pi_2=0$  and therefore $\Pi_0=1$. 
We also note that there exists a modified Ansatz 
for $\Pi_2$ \cite{HB3}, referring to the eigenstates of $\rho_1$ and already taking into 
account that the failure probability 
is to be made as small as possible. This Ansatz implicitly 
contains one of the reduction theorems derived in \cite{raynal}. For the purposes of this 
contribution, however, it is sufficient to rely on the representation in Eq. (\ref{detect1}).

It is our aim  to determine the particular measurement that is optimally 
suited for unambiguous discrimination of the given states.
For this purpose we have to  
insert the general Ansatz for $\Pi_1$ and $\Pi_2$ into the second line of Eq. (\ref{Qfail}) 
and determine the parameters 
$\alpha_{ij}$ and $\beta_{ij}$ that minimize the failure probability 
$Q$ under the constraint 
that the operator $I-\Pi_1-\Pi_2$ is positive. 
So far complete analytical solutions, for arbitrary prior probabilities 
of the two mixed states, have only been found for special cases \cite{rudolph,HB3,BFH}.
All of them are characterized by the fact that the two density operators to 
be discriminated have a certain mutual geometrical structure, where after suitable numbering 
of the eigenvectors we can write for any $l$ and $m$
\begin{eqnarray} 
\label{structure} 
\langle v_i|r_l\rangle  =  \langle v_i|r_i\rangle \delta_{i,l}, \quad
\langle w_j|s_m\rangle & = & \langle w_j|s_j\rangle \delta_{j,m}, 
\end{eqnarray} 
for $i= 1, \ldots, d_{1\perp}$ and $j= 1, \ldots, d_{2\perp}$.
>From this assumption it follows that 
${\rm Tr}({\rho}_1{\Pi}_1)
  = \sum_{i=1}^{d_{1\perp}} \alpha_{ii} r_i |\langle v_i|r_i\rangle|^2$  
and, similarly, ${\rm Tr}({\rho}_2{\Pi}_2)$ also depends on the diagonal elements $\beta_{ii}$ 
only. As a consequence, the 
states $\{|v_i\rangle\}$ and $\{|w_j\rangle\}$ are eigenstates of the 
optimum detection operators $\Pi_1$ and $\Pi_2$, respectively,
and the solution of the optimization problem therefore is facilitated.  

To elucidate our approach, let us treat a simple analytically solvable 
example in which we want to optimally discriminate 
two density operators with $d_1=2$ and 
$d_2=3$. We assume that their supports have a common subspace and 
jointly span a four-dimensional Hilbert space. 
In particular, in terms of the basis vectors $|u_i\rangle$ 
$(i=1,\ldots 4)$ with $\langle u_i|u_j\rangle = \delta_{i,j}$,
we choose  
\begin{eqnarray} 
\label{rho-ex}
\rho_1  = \frac{|u_2\rangle \langle u_2|}{2}  
 + \frac{(|u_3\rangle - |u_4\rangle) 
 (\langle u_3| - \langle u_4|)}{4},
\quad 
\rho_2 =  
  \sum_{i=1}^3  \frac{|u_i\rangle \langle u_i|}{3}. 
\end{eqnarray}
Although for solving the problem one could apply 
one of the reduction theorems derived in \cite{raynal}, 
here we proceed in a direct way.
We find immediately that in Eq. (\ref{12}) $d_{1\perp}=1$ with 
$|v_1\rangle = |u_4\rangle$, while $d_{2\perp}=2$  
with
$|w_1\rangle = |u_1\rangle$ and 
$|w_2\rangle = (|u_3\rangle + |u_4\rangle)/\sqrt{2}$.
Making use of Eqs. (\ref{detect1}) and (\ref{Qfail}) the 
failure probability can be written as  
$Q = 1- \eta_1\alpha_{11}/4 - \eta_2 (2\beta_{11} + \beta_{22})/6$.
In order to minimize $Q$ under the constraint that $\Pi_0$ is positive, we put $\beta_{12}=0$.  
The four eigenvalues of $\Pi_0$ are then found to be equal to 
$1$, $1-\beta_{11}$, and 
$ 2-\beta_{22}-\alpha_{11} \pm \sqrt{\alpha_{11}^2+\beta_{22}^2}$.
They all are positive provided that $\beta_{11}\leq 1$ and 
$\beta_{22}\leq(2-2\alpha_{11})(2-\alpha_{11})$. For making $Q$ 
as small as possible we choose the equality signs. 
After substituting these expressions, we minimize the resulting function 
$Q(\alpha_{11})$, taking into account that $0\leq \alpha_{11} \leq 1$.  
 We find that $Q$ takes its minimum when 
$\alpha_{11}= 0$ if $3\eta_1 \leq \eta_2 $, 
$\alpha_{11}=2(1-\sqrt{\eta_2/3\eta_1})$ 
         if $\eta_2\leq 3\eta_1 \leq 4\eta_2$, and       
$\alpha_{11}=1$ if $3\eta_1 \geq 4\eta_2$.       
This yields the minimum failure probability
\begin{equation}
\label{alpha}
Q_{min}= \left \{
\begin{array}{ll}
1-\frac{\eta_2}{2}  \;\; &  \mbox {if $ \;\;3\eta_1 \leq \eta_2 $}  \\
\frac{\eta_1}{6}+\frac{1}{3}(1+\sqrt{ 3\eta_1\eta_2})
         \;\; & \mbox{if $\;\;\eta_2\leq 3\eta_1 \leq 4\eta_2$}\\       
 1- \frac{\eta_1}{4}-\frac{\eta_2}{3}  
    \;\; & \mbox{if $\;\;3\eta_1 \geq 4\eta_2$},\\       
\end{array}
\right. 
\nonumber
\end{equation}
where $\eta_2=1-\eta_1$. In the inner parameter region the 
optimum measurement is a 
generalized measurement, while in the two outer regions it is 
a von Neumann measurement, where the detection operators are projectors.

\section{General inequalities for the failure probability} 

Although for discriminating two completely arbitrary mixed states
a closed expression for the minimum achievable  
failure probability is lacking, lower bounds can be derived for its value,
in dependence on the prior probabilities of the  states.
For this purpose we first observe that by using the relation between the 
arithmetic and the geometric mean  
as well as the  
Cauchy-Schwarz-inequality \cite{nielsen}   
we obtain from the first line of Eq. (\ref{Qfail}) that \cite{feng}    
\begin{eqnarray}
Q    & \geq &  2\sqrt{\eta_1 \eta_2 
           \rm Tr(\rho_1\Pi_0)\rm Tr(\rho_2\Pi_0)}\nonumber\\  
  &\geq & 2\sqrt{\eta_1 \eta_2}\,  
 {\rm Max}_U\, |{\rm Tr}(U\sqrt{\rho_1}\Pi_0\sqrt{\rho_2})|, 
\label{Q1}
\end{eqnarray}
where $U$ describes an arbitrary unitary transformation. Using $\Pi_0=
I - \Pi_1 - \Pi_2$, as well as the condition
for error-free discrimination, Eq. (\ref{Pi1}),
we arrive at the inequality
\begin{equation} 
\label{Q2}    
Q   \geq  2\sqrt{\eta_1 \eta_2}  
 {\rm Max}_U\,  
 |{\rm Tr}(U\sqrt{\rho_1}\sqrt{\rho_2})|   
  =  2\sqrt{\eta_1 \eta_2}\; F(\rho_1,\rho_2),  
\end{equation} 
where $F={\rm Tr}\,[\left(\sqrt{{\rho}_2}\; {\rho}_1  
\sqrt{{\rho}_2}\right)^{1/2}]$ is the fidelity \cite{nielsen}. 

We can specify the lower bound for the failure probability 
still further when we investigate
the conditions that must be fulfilled when the bound 
$2\sqrt{\eta_1 \eta_2}\; F$ is saturated, i. e. 
when the equality signs hold in Eqs. (\ref{Q1}) and (\ref{Q2}). 
Obviously this is  true if and only if
\begin{equation}
\eta_1 \rm Tr(\rho_1\Pi_0)= \eta_2 \rm Tr(\rho_2\Pi_0) = \sqrt{\eta_1 \eta_2}\; F. 
\label{cond}
\end{equation}
At this point it is important to note that 
the possible values of ${\rm Tr}(\rho_1\Pi_0)$ and ${\rm Tr}(\rho_2\Pi_0)$
are restricted which is a consequence of the structure of the 
detection operators $\Pi_1$ and $\Pi_2$. 
Since the detection operators determine probabilities and, therefore, their eigenvalues 
are between 0 and 1, we get from Eqs. (\ref{detect1}) and (\ref{12}) the relation 
\begin{equation} 
\label{norm} 
0\leq {\rm Tr}(\rho_1\Pi_1)\leq {\rm Tr}(P_{1\perp}\rho_1)= 1 - {\rm Tr}(P_{2}\rho_1). 
\end{equation} 
Taking into account that ${\rm Tr}(\rho_1\Pi_1) = 1 - {\rm Tr}(\rho_1\Pi_0)$  
because of Eqs. (\ref{Pi}) and (\ref{Pi1}), and making use of the symmetry
with respect to interchanging the indices 1 and 2, 
we arrive at the basic inequalities \cite{HB3}
\begin{equation} 
\label{ineq} 
{\rm Tr}(\rho_1\Pi_0)\geq {\rm Tr}(P_{2}\rho_1), \qquad
{\rm Tr}(\rho_2\Pi_0)\geq {\rm Tr}(P_{1}\rho_2). 
\end{equation} 
Obviously due to the requirement that the discrimination be eror-free, 
expressed by Eq. (\ref{Pi1}),
the two state-selective failure probabilities ${\rm Tr}(\rho_1\Pi_0)$ 
and ${\rm Tr}(\rho_2\Pi_0)$ each exceed a 
certain minimum which is larger than zero unless the supports of the density 
operators to be discriminated are orthogonal. 
Combining Eqs. (\ref{cond}) and (\ref{ineq}) we find that the lower bound can only be saturated when 
${\rm Tr}(P_2 \rho_1)/{F} \leq \sqrt{{\eta_2}/{\eta_1}}     
         \leq {F}/{{\rm Tr}(P_1 \rho_2)}$ \cite{HB3}.
 From these considerations it is easy to   
obtain the general inequalities for the failure probability \cite{raynal1}  
\begin{equation}
\label{bound}
Q\geq \left \{ \begin{array}{ll}
\frac{\eta_1 F^2}{{\rm Tr}(P_1 \rho_2)} + \eta_1 {\rm Tr}(P_1 \rho_2) 
& \mbox{if $ \sqrt{\frac{\eta_2}{\eta_1}}     
         \geq \frac{F}{{\rm Tr}(P_1 \rho_2)} $}\\
2\sqrt{\eta_1 \eta_2}F\;\; 
& \mbox{if $ \frac{{\rm Tr}(P_2 \rho_1)}{F} \leq \sqrt{\frac{\eta_2}{\eta_1}}     
         \leq \frac{F}{{\rm Tr}(P_1 \rho_2)} $}\\
\frac{\eta_2 F^2}{{\rm Tr}(P_2 \rho_1)} + \eta_2 {\rm Tr}(P_2 \rho_1) 
& \mbox{if $ \sqrt{\frac{\eta_2}{\eta_1}}     
         \leq \frac{{\rm Tr}(P_2 \rho_1)}{F} $},\\
 \end{array}
\right. 
\end{equation}    
where the conditions on the right-hand side still can be modified taking into account that $\eta_2=1-\eta_1$.

Clearly, the necessary condition
for the saturation of the bound $Q=2\sqrt{\eta_1 \eta_2}F$ can 
be only fulfilled, for certain values of the prior probabilties, if   
\begin{equation}
{\rm Tr}(P_2 \rho_1) {\rm Tr}(P_1 \rho_2)\leq F^2.
\label{cond1}
\end{equation}
When at least one of the states to be discriminated is pure, there exists always a parameter
interval for the prior probabilites where the failure probability reaches 
the fidelity bound $Q=2\sqrt{\eta_1 \eta_2}F$, as can be seen from the exact solution \cite{BHH,BHH1}.
For two arbitrary mixed states, however, this statement does not hold true \cite{HB3}. 
In fact,
the condition expressed by Eq. (\ref{cond1}) can be violated even when the supports of the two density operators
do not have a common subspace. 
 \begin{figure}[ht]
\begin{center}
\epsfig{file=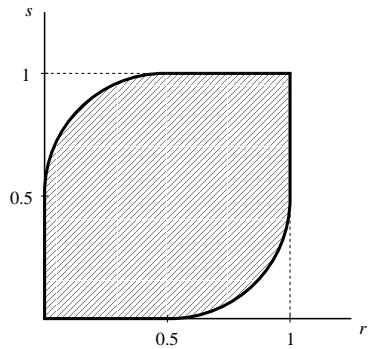, height=4.5cm}
\caption
{The parameter space of the two special density operators from Eq. (\ref{rho-s}). 
In the shaded area the inequality (\ref{cond1}) is satisfied, $F^{2} 
\geq \frac{1}{4}$, and the white corners represent the part of the parameter 
space where it is violated, $F^{2} < \frac{1}{4}$. The boundaries of the 
shaded area are given by $s=0$ for $0 \leq r \leq 1/2$, by 
$(r-1/2)^{2}+(s-1/2)^{2}=1/4$ for $1/2 \leq r \leq 1, \ 0 \leq s \leq 1/2$ 
and for $0 \leq r \leq 1/2, 1/2 \leq s \leq 1$, and by $s=1$ for $1/2 \leq r 
\leq 1$.} 
\end{center}
\end{figure}

As an illustration, we specify Eq. (\ref{rho}) and consider the density operators 
\begin{eqnarray}
\rho_1  =  r |r_1\rangle + (1-r) |r_2\rangle,  \quad
\rho_2  =  s |s_1\rangle + (1-s) |s_2\rangle,
\label{rho-s}
\end{eqnarray}
where $\langle r_{i}|s_{j}\rangle = \frac{1}{\sqrt{2}}\delta_{i,j}$.
It is easy to check that in this case  Eqs. (\ref{structure}) hold.
A short calculation shows that  $F=\frac{1}{\sqrt{2}}[\sqrt{rs} + 
\sqrt{(1-r)(1-s)}]$ and ${\rm Tr}(P_{1}\rho_2)={\rm Tr}(P_{2}\rho_1)=1/2$. 
As becomes obvious from Fig. 1, there exists a parameter space for $(r,s)$ where  
the necessary condition (\ref{cond1}) is violated and the fidelity bound 
cannot be reached for any value of the prior probabilities. 
We still mention that for $r=s$ the exact solution for the minimum failure probability
in our example is given by Eq. (\ref{bound}) when on the left-hand side the equality sign holds, as has been shown in \cite{HB3}. 
In this case again in the intermediate parameter region, where the fidelity bound is reached,  
the optimum measurement is a generalized measurement, while in the outer two regions it is a von Neumann measurement.

\section{Conclusions}
In this contribution we briefly rederived some of our recent results \cite{HB3}  on the problem of 
optimum unambiguous discrimination of two mixed states, and we discussed 
two illustrative special cases.
It is worth noting that even for the 
lowest-dimensional non-specialized case, where two completely arbitrary density operators 
of rank two have to be distinguished, 
so far there does not exist a general analytical solution.
 
\ack
J. B. acknowledges useful discussions with 
E.~Feldman and M. Hillery.


\section*{References}


\begin{thebibliography}{99}   
   
 
\bibitem{springer} for a recent review see, e. g.,  
J. A Bergou, U. Herzog, and M. Hillery, Lect. Notes Phys. 649, 
417-465 (Springer, Berlin, 2004). 

\bibitem{helstrom}
C. W. Helstrom, {\it Quantum Detection and Estimation Theory}  
    (Academic Press, New York, 1976).  


\bibitem{ivan} I. D. Ivanovic, Phys. Lett. {\bf A123}, 257   
  (1987),    

\bibitem{dieks}
D. Dieks, Phys. Lett. {\bf A126}, 303 (1988).

\bibitem{peres}  
 A. Peres, Phys. Lett. {\bf A128}, 19 (1988).  

\bibitem{jaeger} G.\ Jaeger and A.\ Shimony, Phys.\ Lett.\ {\bf A197},   
  83 (1995).     
   
\bibitem{SBH} 
 Y. Sun, J. A. Bergou, and M. Hillery, Phys. Rev. A {\bf      
  66}, 032315 (2002).           
 
\bibitem{BHH}  
J. A Bergou, U. Herzog, and M. Hillery,    
Phys. Rev. Lett. {\bf 90}, 257901 (2003), 

\bibitem{BHH1} J. A Bergou, U. Herzog, and M. Hillery,    
Phys. Rev. A {\bf 71}, 042314 (2005).    

\bibitem{rudolph} T. Rudolph, R. W. Spekkens, and P. S. Turner,    
  Phys. Rev. A {\bf 68}, 010301(R) (2003).   
   
\bibitem{raynal} Ph. Raynal, N. L\"utkenhaus, and S. van Enk,   
  Phys. Rev. A {\bf 68}, 022308 (2003).   
  
\bibitem{HB2} U. Herzog and J. A. Bergou, Phys. Rev. A {\bf 70},  022302 (2004).    

   
\bibitem{feng} Y. Feng, R. Duan, and M. Ying, Phys. Rev. A {\bf 70},   
12308 (2004).  
   
\bibitem{HB3} U. Herzog and J. A. Bergou, Phys. Rev. A  {\bf 71}, 050301 (R) (2005).  

\bibitem{raynal1} Ph. Raynal and N. L\"utkenhaus, Phys. Rev. A {\bf 72}, 022342 
(2005).

\bibitem{BFH} J. A. Bergou, E. Feldman, and M. Hillery 
(submitted to Phys. Rev. A)

\bibitem{neumark} M.\ A.\ Neumark,    
Izv. Akad. Nauk. SSSR, Ser. Mat. {\bf 4}, 277 (1940).    
   
    
\bibitem{nielsen}  M. A. Nielsen and I. L. Chuang,    
{\it Quantum Computation and Information} (Cambridge University Press, 2000).   
   
\end{thebibliography}
 \end{document}